\documentclass[11pt,onecolumn,showpacs,superscriptaddress,groupedaddress]{revtex4}

\usepackage{graphicx}  % needed for figures
\usepackage{dcolumn}   % needed for some tables
\usepackage{bm}        % for math
\usepackage{amssymb}   % for math
\usepackage{subfigure}

\begin{document}

\title{On the Tunneling Properties of non-Planar Molecules in a Gas Medium}

\author{Mohammad Bahrami}
\email{mbahrami@ictp.it} \affiliation{Department of Chemistry, Sharif University of Technology, P.O.Box 11365-9516, Tehran, Iran } \affiliation{The Abdus Salam ICTP, Strada Costiera 11, 34151 Trieste, Italy}

\author{Angelo Bassi}
\email{bassi@ts.infn.it} \affiliation{Department of Physics, University of
Trieste, Strada Costiera 11, 34151 Trieste, Italy} \affiliation{Istituto
Nazionale di Fisica Nucleare, Trieste Section, Via Valerio 2, 34127 Trieste,
Italy}

\begin{abstract}
We propose a simple, general, and accurate formula for analyzing the tunneling between classical configurations of a non-planar molecule in a gas medium, as a function of the thermodynamic parameters of the gas. We apply it to two interesting cases: i) The shift to zero frequency of the inversion line of ammonia, upon increase of the pressure of the gas; ii) The destruction of the coherent tunneling of D$_2$S$_2$ molecules in a gas of He. In both cases, we compare our analysis with previous theoretical and experimental results.

\end{abstract}


\maketitle

\section{Introduction.}
Quantum tunneling has always been one of the most fascinating---and counterintuitive---manifestation of quantum phenomena in nature~\cite{Mer,LW}. Among the most well-known examples of this kind, are non-planar molecules, in which an atom or a group of atoms oscillates between the two wells of the potential energy surface, also at low temperatures~\cite{Town,hund,Mer}. There has been a great amount of interest in the study of coherent tunneling in molecular systems, and how and under which physical conditions this coherence may be suppressed due to environmental interactions~\cite{HS,SH,Sim,Pf,JZ,Jona,Ver,Wig,TH,AK,Ber,TT,deco,BP}.

The relevant property of the molecule can be effectively modeled by a particle of mass $M$ moving in a double-well potential $V(q)$ with two minima at $q=\pm q_{0}/2$; here $q$ is a generalized inversion coordinate. The minima, associated with the two localized (say, chiral) states of the molecule, are separated by a barrier $V_{0}$. If we denote the small-amplitude vibration in either well by $\omega_{0}$, then in the limit
$ V_{0} \gg \hbar\omega_{0} \gg k_{B}T, $ (where $T$ is the temperature of the bath and $k_{B}$ is Boltzmann constant), the state of the system is effectively confined in the two-dimensional Hilbert space spanned by the two lowest eigenstates.

As custom, we denote by $|+\rangle$ and $|-\rangle$ the ground and first excited state of the relevant part of the Hamiltonian, separated by the energy difference $\Delta E = \hbar\omega_{x}$. They are delocalized states, since they must be also eigenstates of the parity operator. The localized---or chiral---states, are the combinations:
\begin{equation}
\label{tab:eq1}
|\text{L}\rangle = \frac{1}{\sqrt{2}} [ |+\rangle + |-\rangle ], \qquad
|\text{R}\rangle = \frac{1}{\sqrt{2}} [ |+\rangle - |-\rangle ].
\end{equation}
In this two--dimensional approximation, the Hamiltonian takes the form: $\hat{H}=-\omega_{x}\hat{\sigma}_{x}/2,$ where  $\hat{\sigma}_{x}$ has $|\pm\rangle$ as eigenstates. The position operator $\hat{q}$ reduces to $(q_0/2)\hat{\sigma}_{z}$, which has the localized states $|\text{L}\rangle, \: |\text{R}\rangle$ as eigenstates.

\section{The role of decoherence.} 
A very interesting problem is the behavior of a non-planar molecule in a gas, and how the tunneling between different classical configurations depends on the thermodynamic variables of the gas. Several approaches have been developed, for analyzing the system-bath interaction. One of the best understood ones is decoherence~\cite{LW,deco,BP}, and in particular collisional decoherence~\cite{JZ,Dios,VH,Adl}, which is ubiquitous in quantum physics. The application of decoherence to the study of the coherent tunneling of a molecule in a gas has been first done by~\cite{HS,SH}. Different alternative approaches to the problem include the mean field theory~\cite{Jona}, or various modeling of the dephasing process induced by intermolecular
interactions~\cite{Ver}, by photons~\cite{Pf}, or by phonons~\cite{SH}. However, decoherence seems to cover the widest ranges of physical situations.

When analyzing decoherence phenoma, the starting point is the Lindblad-type master equation~\cite{lind}, which in the two-dimensional approximation reads:
\begin{equation}
\label{tab:eq2}
\frac{\partial \hat{\rho}}{\partial t} =
-\frac{i\omega_{x}}{2}\left[\hat{\sigma}_{x},\hat{\rho}\right]
-\frac{\lambda}{2}\left(\hat{\rho}-\hat{\sigma}_{z}\hat{\rho}\hat{\sigma}_{z}\right),
\end{equation}
where $\hat{\rho}$ is the density matrix of the relevant molecule, and $\lambda$ the decoherence rate. The task of (collisional) decoherence is to give a microscopic expression for $\lambda$.
% This type of microscopic derivation has been first performed in \cite{TH}, where the decoherence effect on $\text{D}_2\text{S}_2$ molecules by He atoms as a background gas has been computed. The calculation is highly complex, time-consuming and dependent on the details of internal-rotational states of system under study.
In this article we propose a general master formula for computing the decoherence rate $\lambda$ for non-planar molecules. The advantage of our formula---differently from previous approaches---is that it can be easily applied to many different molecules in a large variety of background gases. As we will show, even in its simplest form, it gives results which are in excellent agreement with known data.

\subsection{A master formula for $\lambda$.} 
The key, and simple, observation is that energy eigenstates are superpositions in {\it space} of one or more atoms of the molecule, and for this reason they decohere upon interaction with an environment. Therefore we may apply the theory of quantum Brownian motion~\cite{VH} to such atoms, in order to compute the loss of coherence due to the scattering with the bath particles. In the dilute gas limit, and when the scattering is recoil-free, the decoherence master equation takes the form \cite{Dios,VH,Adl}:
\begin{equation}
\label{tab:eq3}
\frac{\partial \left\langle \textbf{x}|\hat{\rho}|\textbf{x}'\right\rangle}{\partial t} =
-\frac{i}{\hbar}\langle \textbf{x}|[\hat{H} ,\hat{\rho}]|\textbf{x}'\rangle
-F\left(\textbf{x}-\textbf{x}'\right)\left\langle \textbf{x}|\hat{\rho}|\textbf{x}'\right\rangle
\end{equation}
with $\hat{H}$ the Hamiltonian of the isolated Brownian particle, $\left|\textbf{x}\right\rangle, \left|\textbf{x}'\right\rangle$ its position eigenstates, and:
\begin{equation}
F\left(\textbf{x}\right)\; = \; n
\int_{0}^{\infty} dp \: \nu\left(p\right) \frac{p}{m} \: \sigma\left(p,\textbf{x}\right)
\end{equation}
where
$n$ is the density of bath particles,
$m$ the reduced mass of bath particles,
$p$ the bath particle momentum,
$\nu\left(p\right)$ the distribution of such momenta, and:
\begin{equation} \label{eq:sigma}
\sigma\left(p,\textbf{x}\right)=\int \frac{d\hat{\textbf{n}}_{1}d\hat{\textbf{n}}_{2}}{4\pi}
\left(1-e^{ip\left(\hat{\textbf{n}}_{1}-\hat{\textbf{n}}_{2}\right)\cdot\textbf{x}/\hbar}\right)
\left|f\left(p,\theta\right)\right|^{2}
\end{equation}
is the decoherence cross section; $f\left(p,\theta\right)$ is the scattering amplitude, and $\theta$ the scattering angle between directions $\hat{\textbf{n}}_{1}$ and $\hat{\textbf{n}}_{2}$.

In the limit where the two-dimensional approximation is valid, the spatial coherence is between the two minima of the double--well potential, which are separated by a distance $\left|\textbf{x}-\textbf{x}'\right|=q_{0}$. We assume that, for all important bath particles' momenta $p$, the decoherence cross section $\sigma\left(p,q_{0}\right)$ does not change appreciably. We can then write:
\begin{equation}
\lambda \; \equiv \; F(q_0) \; \simeq \; \frac{n \bar{p}}{m}\, \sigma\left(\bar{p},q_{0}\right),
\end{equation}
where $\bar{p} = \sqrt{2mk_B T}$ is the most probable momentum, for an ideal gas at equilibrium at temperature $T$. For isotropic scattering, we have~\cite{VH}:
\begin{equation}
\label{tab:eq4}
\sigma\left(\bar{p},q_{0}\right)=2\pi \int^{+1}_{-1} d\left(\cos \theta\right)
\left(1-\frac{\sin \Theta}{\Theta}\right)\left|f\left(\bar{p},\theta\right)\right|^{2}
\end{equation}
with $\Theta=(2\bar{p}q_{0}/\hbar)\sin\left(\theta/2\right)$. Eq.~(\ref{tab:eq4}) can be easily computed in two limiting cases: the low temperature limit $\bar{p} q_{0} \ll \hbar$, i.e. $T \ll \hbar^2/2mk_Bq_0^2$; and the high temperature limit $\bar{p} q_{0} \gg \hbar$, i.e. $T \gg \hbar^2/2mk_Bq_0^2$.

For values of $\omega_{0}$ in the infra--red range ($10^{13}$--$10^{14}\:$Hz), which is the typical range of molecular vibrations, $q_{0}$ of the order of few Angstrom, and $m$ of $2$--$30\:\text{gr/mol}$ for masses of typical background gases (e.g., from H$_2$ to Air), one has: $\hbar\omega_{0}/k_{B} \approx 500$--$5000\:\text{K}$ and $\hbar^{2}/2mk_{B}q_{0}^{2} \approx 1$--$30\:\text{K}$. This means that both the two dimensional approximation and the high temperature limit are valid in most interesting physical situations. In this case, the variable $\Theta$ is large, $\sin\Theta/\Theta \simeq 0$, thus: $\sigma\left(\bar{p},q_{0}\right)=\sigma_{\text{\tiny TOT}}\left(\bar{p}\right)=2\pi\int d\left(\cos\theta\right) \left|f\left(\bar{p},\theta\right)\right|^{2}$, which is the total cross-section.

If we consider the scattering interaction as given by a central potential and use the phase-shift method \cite{Mes,Joach}, the scattering amplitude $f\left(\bar{p},\theta\right)$ can be expressed as:
$f\left(\bar{p},\theta\right)=
(\hbar/\bar{p})\sum_{\ell=0}^{\infty}\left(2\ell+1\right)e^{i\delta_{\ell}}
\sin{\delta_{\ell}}\:P_{\ell}\left(\cos{\theta}\right)$,
with $P_{\ell}\left(\cos{\theta}\right)$ the Legendre polynomials, and $\delta_{\ell}$ the phase shifts. The phase shifts have suitable integral representations depending on the scattering potential.
Collecting the above results, we can write:
\begin{equation}
\label{tab:eq5}
\lambda \; = \; \frac{n\bar{p}}{m}\, \sigma_{\text{\tiny TOT}}(\bar{p}) \; = \;
\frac{4\pi P\hbar^{2}}{\sqrt{2}(mk_B T)^{3/2}}
\sum_{\ell=0}^{\infty}\left(2\ell+1\right)\sin^{2}{\delta_{\ell}}.
\end{equation}
where we used $n = P/k_B T$.

Eq.~(\ref{tab:eq5}) and, with it, decoherence theory, is very powerful: it tells that the effect of a gas medium on the tunneling properties of a molecule can be reduced to the two-body interaction (scattering) between the relevant molecule and a single gas molecule. The gas as a whole enters only though its thermodynamics parameters. Eq.~(\ref{tab:eq5}) is also very general, because can be immediately applied to any situation, where a non-planar molecule is immersed in a gas medium (within the limits of validity of the general approach, previoulsy discussed). Eq.~(\ref{tab:eq5}) is also independent from the internal-rotational states of the non-planar molecule.

Computing the phase shifts $\delta_{\ell}$ can prove quite hard. One can provide a handy expression in the hard-sphere limit:  $\sin^2\delta_{\ell}=j^2_{\ell}\left(\bar{p}a/\hbar\right)/
[j^2_{\ell}\left(\bar{p}a/\hbar\right)+n^2_{\ell}\left(\bar{p}a/\hbar\right)]$ with $a$ the distance of closest approach of the centers of the two colliding spheres, $j_{\ell}\left(x\right)$ the spherical Bessel functions and $n_{\ell}\left(x\right)$ the spherical Neumann functions~\cite{Mes,Joach}. The hard-sphere limit is the zero-th order approximation one can make. However it turns out to be very good, because intermolecular interactions typically have a finite effective range, and the main contribution to $\lambda$ comes essentially from scattering in this effective region. Considering the effective range as that of a hard-sphere, we can find a very simple and general expression for $\lambda$. In fact, as we will show in the following examples, the hard-sphere limit provides a very good description of decoherence effects in the tunneling properties of a molecule. It also offers a geometric, and very intuitive, way to understand the decoherence mechanism, reducing it to the bouncing of two spheres. In this way, one can understand why, in the calculations, one has to include a steric factor $\gamma$, which takes into account only those directions of approach of a gas molecule, which can contribute to decoherence: an atom attached to a molecule can be ``hit'' by a bath particle only from the directions which are not hindered by the rest of the molecule.
We now apply this setup to two interesting physical situations.

\section{Shift in inversion-line frequency of ammonia gas.} 
\begin{figure}
\begin{center}
{\includegraphics[scale=1]{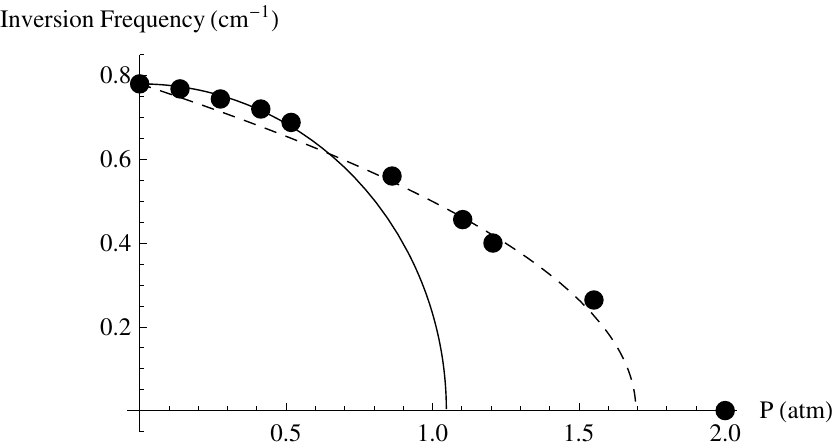}}
\caption{Inversion frequency of ammonia, as a function of the gas pressure $P$. The dots are experimental data taken from~\cite{BL}; the solid line refers to Eq.~(\ref{eq:decrate}), with $P_{\text{\tiny cr}} = 1.05\,$atm; the dashed line refers to Eq.~(\ref{eq:decratejl}) of~\cite{Jona}, with $P_{\text{\tiny cr}} = 1.695\,$atm. In both cases, $\omega_x$ (frequency at $P=0$) has been taken equal to 0.78 cm$^{-1}$, as reported in~\cite{BL}.}
\label{trid}
\end{center}
\end{figure}

In~\cite{BL}, the inversion-line frequency of ammonia has been experimentally measured as a function of the gas pressure $P$. The frequency is found to decrease from 0.78 cm$^{-1}$ at $P = 0$, to zero above $P = 2$ atm (See Fig.~\ref{trid}). This phenomenon was first explained qualitatively in \cite{and,marg}; to our knowledge, the first complete theoretical analysis is provided in \cite{Jona}, based on the mean field theory. The inversion frequency turns out to be:
\begin{equation} \label{eq:decratejl}
\bar{\omega}_x \; = \; \omega_x \, \sqrt{1 - \frac{P}{P_{\text{\tiny cr}}}},
\end{equation}
where $P_{\text{\tiny cr}} = 1.695\,\text{atm}$ is the critical pressure derived theoretically. The theoretical formula, reported in Fig.~\ref{trid}, is in very good agreement with the experimental data.

The mean field theory provides a `static' approach to the problem. Decoherence theory provides a `dynamical' approach, and its predictions partly differ, as we now show. This comes through the explicit solution of Eq.~(\ref{tab:eq2}). By writing $\hat{\rho}=(1/2)\left(I+P_{x}\hat{\sigma}_{x}+P_{y}\hat{\sigma}_{y}+P_{z}\hat{\sigma}_{z}\right)$, all information is contained in the Bloch vector ${\bf P} \equiv (P_x, P_y, P_z)$. The experimental results obtained by cavity-resonator technique are connected  to the relative occupation probability
$P_{z}\left(t\right)$ \cite{BL}, whose time evolution, in the under--damped case, reads \cite{Bas}:
\begin{eqnarray}
P_{z}\left(t\right) & = & \frac{e^{-t\lambda/2}}{2\bar{\omega}_x}
[
2P_z(0)\bar{\omega_x}\cos \bar{\omega}_xt
\\
\nonumber
&+&
(P_z(0)\lambda+2P_y(0)\omega_x)\sin \bar{\omega}_xt
],
\end{eqnarray}
where the characteristic inversion frequency, $\bar{\omega}_x$, is:
\begin{equation} \label{eq:decrate}
\bar{\omega}_x \; = \; \omega_x \, \sqrt{1 - \left(\frac{P}{P_{\text{\tiny cr}}}\right)^2},
\end{equation}
and $P_{\text{\tiny cr}} = 2 P \omega_x / \lambda$. Since the experiment in~\cite{BL} has been performed at room temperature, the high temperature limit holds and, according to Eq.~(\ref{tab:eq5}), we can write the critical pressure as follows:
\begin{equation} \label{eq:pcr}
P_{\text{\tiny cr}} = \frac{\sqrt{2\omega_x^2 m k_B T}}{\sigma_{\text{\tiny TOT}}(\bar{p})},
\end{equation}
As expected, Eq.~(\ref{eq:decrate}) predicts a decrease of the inversion frequency, due to the interaction of each molecule with the rest of the gas. Moreover, Eq.~(\ref{eq:pcr}) provides a microscopic expression for $P_{\text{\tiny cr}}$, without free parameters.

A good estimate for the total cross section $\sigma_{\text{\tiny TOT}}(\bar{p})$ can be provided in the hard-sphere limit, previously discussed. The hard-sphere radius is equal to twice the effective hard-sphere radius of ammonia ($a=4.38$ \AA) \cite{Mor}.
In computing the total cross section, the relevant contribution comes from those terms with $\ell \leq \bar{p} a / \hbar \simeq 22$ \cite{Mes,Joach}; this gives $\sigma_{\text{\tiny TOT}} = 481a_{\text{\tiny Bohr}}^{2}$ (where $a_{\text{\tiny Bohr}}$ is Bohr radius). In the calculations, we have chosen the masses as follows. According to the double-well description, the mass $M$ of the Brownian particle corresponds to the reduced mass $M =\frac{3m_{\text{\tiny H}} m_{\text{\tiny N}}}{3m_{\text{\tiny H}}+m_{\text{\tiny N}}}$~\cite{herz} with $m_{\text{\tiny H}}$ mass of Hydrogen and $m_{\text{\tiny N}}$ mass of Nitrogen. The mass $m$ entering the calculation of the total cross section and of the decoherence rate refers to the reduced mass of the Brownian and bath particle ($m_*$ in Eq.~(2.3) of ~\cite{VH}): $m=\frac{M \,m_{\text{\tiny NH}_3}}{M+m_{\text{\tiny NH}_3}}$, with $m_{\text{\tiny NH}_3}$ the mass of ammonia. We have to multiply the cross section by a steric factor $\gamma \simeq 0.85$, since not all collisions of a bath particle with the Brownian particle result to decoherence (see Fig.~2(a)). By taking $\omega_x = 0.78$ cm$^{-1}$ and $T = 300$K~\cite{BL}, Eq.~(\ref{eq:pcr}) gives: $P_{\text{\tiny cr}}=1.05\,$atm.

As shown in Fig.~\ref{trid}, Eq.~(\ref{eq:decrate}) matches well the experimental data at low pressures, but clearly fails at high pressures. This does not come as a surprise, since the microscopic model used to compute $\lambda$ is valid only in the diluted gas case. As explained in \cite{Town}, in the case of ammonia, the region $P\geq1\,\text{atm}$, where many-body collisions become important, is far from the dilute gas limit.

It is important to note that Eq.~(\ref{eq:decrate}) and Eq.~(\ref{eq:decratejl}) predict a different behavior at very low pressures. While, according to Eq.~(\ref{eq:decrate}), the inversion frequency $\bar{\omega}_x$ approaches $\omega_x$ with an horizontal slope for $P \rightarrow 0$, according to Eq.~(\ref{eq:decratejl}) the slope is negative. The experimental data seem to suggest that the slope gradually becomes horizontal, confirming our prediction.

There is also a strong theoretical reason why it should be so. Eq.~(\ref{eq:decrate}) is a direct consequence of Eq.~(\ref{tab:eq2}), i.e. of a Lindblad-type master equation, which is the result of very general mathematical requirements, such as the quantum dynamical semigroup (QDS) structure, and complete positivity~\cite{lind}.  Thus, it would be interesting to perform the experiment again, to assess the low pressure behavior of the inversion-line frequency of ammonia.

\section{Effect of decoherence on tunneling dynamics of $\text{D}_{2}\text{S}_{2}$ molecule.} In~\cite{TH} the decoherence rate $\lambda$ of a bath of Helium (He) on a $\text{D}_{2}\text{S}_{2}$ molecule is studied with heavy numerical machinery. Actually, its value is not explicitly given, but can be estimated from Eq.~(2) of the paper. We assume that the decoherence cross section $\eta_{tot}$ (notation of~\cite{TH}) is constant over significant bath particles' momenta and, according to Fig. (1) of the paper, equal to $103a_{\text{\tiny Bohr}}^{2}$ for the considered temperature $T=300\:\text{K}$. Then Eq.~(2) of the paper gives $\lambda = 215.5\,$Hz.
\begin{figure}
\centering
\subfigure[]{\label{fig:NH3col}\includegraphics[scale=.5]{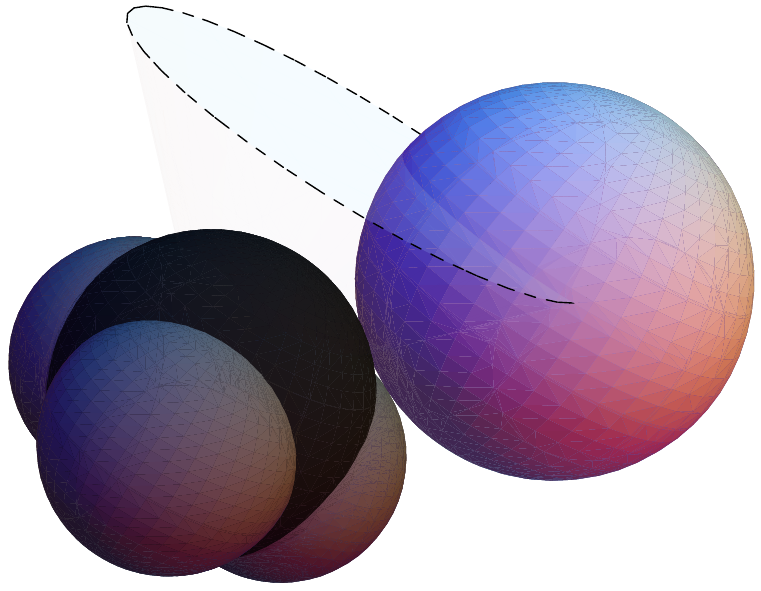}}
\subfigure[]{\label{fig:D2S2col}\includegraphics[scale=.5]{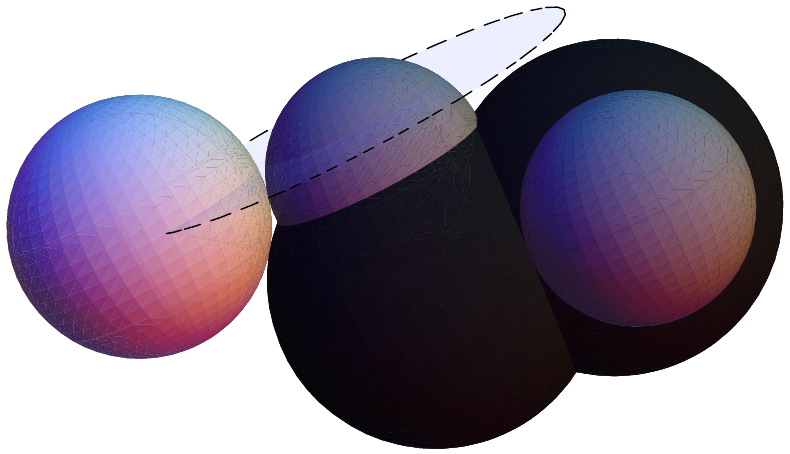}}
\caption{(a) Collision between an NH$_3$ Brownian molecule (left) and a NH$_3$ bath particle (right); we simplify the description by considering the bath particles as spheres. When a bath particle collides with the Brownian particle within the solid angle $\Omega$ depicted in the picture, there is no significant decoherence effect, because the H atoms (who effectively participate to the superposition) are not hit. Accordingly, the geometric cross section must be multiplied by a steric factor $\gamma = (4 \pi - \Omega)/ 4\pi$. This takes into account only those collisions which give rise to decoherence. With simple geometric considerations, one can estimate the angle $\theta$ which identifies $\Omega$ as: $\theta \simeq 1.088$. Then $\Omega=2\pi\int^{\theta}_{0}\sin x dx \simeq 1.875$, and $\gamma \simeq 0.851$
(b) Collision of a Brownian D$_2$S$_2$ molecule (right) with a bath He particle (left). In this case, only the D atoms participate to the superposition. Each D atom is hit by a bath particle when this arrives within the solid angle $\Omega$ shown in the picture. Simple geometric considerations lead to the result: $\Omega \simeq 4.943$. The steric factor for each D atom then is $\gamma = \Omega/4 \pi \simeq 0.393$.
}
\label{fig:2}
\end{figure}

We now apply our framework to give an independent estimate for $\lambda$, using Eq.~(\ref{tab:eq5}).
The Brownian particle has the reduced mass $M=\frac{m_{\text{\tiny D}}\,m_{\text{\tiny S}}}{m_{\text{\tiny D}}+m_{\text{\tiny S}}}$ with $m_{\text{\tiny S}}$ the mass of Sulfur and $m_{\text{\tiny D}}$ the mass of Deuterium.
For $m$ we use $m=\frac{M\, m_{\text{\tiny He}}}{M+m_{\text{\tiny He}}}$
with $m_{\text{\tiny He}}$ the mass of He.
In the hard-sphere limit, it is reasonable to consider $a$ as the sum of the hard-sphere radius of He (1.085 \AA) and the van der Waals radius of D (1.2 \AA)~\cite{Mor}.
Under the conditions of $P=1.6\times10^{-3}\:\text{Pa}$ and $T=300\:\text{K}$ used for the experiment proposed in~\cite{TH}, the high temperature limit holds. In computing the total cross section, the relevant contribution comes from those terms with $\ell \leq \bar{p} a / \hbar \simeq 9$ \cite{Mes,Joach}. One obtains $\sigma_{\text{\tiny TOT}}(\bar{p}) = 139.4 a_{\text{\tiny Bohr}}^{2}$. This result has to be multiplied by a factor 2, since there are two D atoms superimposed for each molecule, and also by a steric factor $\gamma \simeq 0.393$, which takes into account that each D atom does not collide with bath particles coming from all directions, due to the presence of the sulfur atoms (see Fig.\ref{fig:D2S2col}). This gives $\sigma_{\text{\tiny TOT}}(\bar{p}) = 109.7a_{\text{\tiny Bohr}}^{2}$ and, according to Eq.~(\ref{tab:eq5}), $\lambda=229.4\:\text{Hz}$. Our result is in very good agreement with the one predicted by~\cite{TH}.

\section{Conclusions.} Our approach to the tunneling properties of non-planar molecules in a gas offers a simple and general formula for computing the decoherence rate, and its dependence on the thermodynamic variables of the gas. It is flexible and applies to all situations where a molecule is immersed in a gas, which is the most common case, when tunneling properties are under study. The only three limitations are: the diluted-gas limit, the two-dimensional approximation and the high-temperature limit, which are satisfied in most physical conditions. The hard sphere limit represents the grossest type of approximation, but as we have seen its predictions are in excellent agreement with previous analyses. And it has the advantage of reducing a complex many-body problem to a simple geometric calculation. This opens the way to easy estimates of decoherence effects also for very complicated molecules in highly structure gases.

{\it Acknowledgments.} M.B. acknowledges partial financial support of National Elite Foundation-Iran, and hospitality from The Abdus Salam ICTP, where this work was carried out. A.B. acknowledges partial financial support from MIUR (PRIN 2008), INFN and COST (MP1006).


\begin{thebibliography}{99}

\bibitem{Mer}
E. Merzbacher: \textit{Phys. Today} {\bf 44} (Aug. 2002).

\bibitem{LW}
A. J. Leggett \textit{et. al.}: Rev. Mod. Phys. {\bf 59}, 1 (1987).
U. Weiss, Quantum Dissipative Systems (World Scientific, Singapore, 2008).

\bibitem{hund}
W. Hund: Z. Phys. \textbf{40}, 742 (1927);
Z. Phys. \textbf{43}, 805 (1927).

\bibitem{Town}
C. H. Townes and A. L. Schawlow, Microwave Spectroscopy, (McGraw-Hill, New York, 1955).

\bibitem{HS}
R. A. Harris, L. Stodolsky: Phys. Lett. B \textbf{78}, 313 (1978);
J. Chem. Phys. \textbf{74}, 2145 (1981);
Phys. Lett. B \textbf{116}, 464 (1982);
J. Chem. Phys. \textbf{78}, 7330 (1983).

\bibitem{SH}
R. Silbey, R. A. Harris: J. Chem. Phys. \textbf{80}, 2615 (1984);
J. Phys. Chem. \textbf{93}, 7062 (1989).

\bibitem{Sim}
M. Simonius: Phys. Rev. \textbf{40}, 980 (1978).

\bibitem{Pf}
P. Pfeifer: Phys. Rev. A. \textbf{26}, 701 (1982).

\bibitem{JZ}
E. Joos and H.D. Zeh: Z. Phys. B \textbf{59}, 223 (1985).

\bibitem{Jona}
G. Jona-Lasinio, C. Presilla, C. Toninelli: Phys. Rev. Lett. \textbf{88}, 123001 (2002).

\bibitem{Ver}
A. Vardi: J. Chem. Phys. \textbf{112}, 8743 (2000).

\bibitem{Wig}
A. Wightman: Il nuovo Cimento \textbf{110} B,751(1995).

\bibitem{TH}
J. Trost, K. Hornberger: Phys. Rev. Lett. \textbf{103}, 023202 (2009).

\bibitem{AK}
V. Athalye, A. Kumar: J. Phys. B: At. Mol. Opt. Phys. {\bf 39}, 2633 (2006).

\bibitem{Ber}
Y. A. Berlin \textit{et. al.}: Z. Phys. D {\bf 37}, 333 (1996).

\bibitem{TT}
K. Tamagakea, M. Tsuboia, K. Takagib and T. Kojima: Jour. Mol. Spec. {\bf 39},  454 (1971).

\bibitem{deco}
D. Giulini, E. Joos, C. Kiefer, J. Kupsch, I.-O. Stamatescu, and H. D. Zeh, Decoherence and the Appearance of a Classical World in Quantum Theory
(Springer, Berlin 1996).

\bibitem{BP}
H.-P. Breuer and F. Petruccione, The Theory of Open Quantum Systems, (Oxford Univ. Press, Oxford, 2002).

\bibitem{Dios}
L. Diosi, Europhys. Lett. {\bf 30}, 63 (1995).

\bibitem{VH}
B. Vacchini and K. Hornberger, Phys. Rep. {\bf 478}, 71 (2009).

\bibitem{Adl}
S. L. Adler, J. Phys. A: Math. Gen. {\bf 39}, 14067 (2006).

\bibitem{lind}
G. Lindblad, Commun. Math. Phys. {\bf 48}, 119 (1976).

\bibitem{Mes}
A. Messiah, Quantum Mechanics, (New York: Wiley 1999), Chapter X.

\bibitem{Joach}
C. J. Joachain, Quantum Collision Theory (Elsevier Science Publisher B.V., Amsterdam, 1975), Chapter 4.

\bibitem{BL} B. Bleaney and J. H. Loubster, Nature (London) 161, 522 (1948); B. Bleaney and J. H. Loubster, Proc. Phys. Soc. London
Sect. A 63, 483 (1950).

\bibitem{and} P.W. Anderson, Phys. Rev. {\bf 75}, 1450 (1949).

\bibitem{marg} H. Margenau, Phys. Rev. {\bf 76}, 1423 (1949).

\bibitem{Bas}
A. Bassi and E. Ippoliti, Phys. Rev. A {\bf 69}, 012105 (2004).

\bibitem{Mor}
R.C. Weast, ed., Handbook of Chemistry and Physics, 64th ed. (CRC
Press, Boca Raton, FL, 1983) p. F43ff.

\bibitem{herz}
G. Herzberg: Molecular Spectra and Molecular Structure. Electronic Spectra and Electronic
Structure of Polyatomic Molecules (Krieger: Malabar, FL, Reprint 1991) Vol. III.


\end{thebibliography}
\end{document}